\input epsf
\newfam\scrfam
\batchmode\font\tenscr=rsfs10 \errorstopmode
\ifx\tenscr\nullfont
        \message{rsfs script font not available. Replacing with calligraphic.}
        
\else   
        \font\sevenscr=rsfs7
        \font\fivescr=rsfs5
        \skewchar\tenscr='177 \skewchar\sevenscr='177 \skewchar\fivescr='177
        \textfont\scrfam=\tenscr \scriptfont\scrfam=\sevenscr
        \scriptscriptfont\scrfam=\fivescr

\fi
\catcode`\@=11
\newfam\frakfam
\batchmode\font\tenfrak=eufm10 \errorstopmode
\ifx\tenfrak\nullfont
        \message{eufm font not available. Replacing with italic.}
        
\else
	
	\font\sevenfrak=eufm7 \font\fivefrak=eufm5
        
	\textfont\frakfam=\tenfrak
	\scriptfont\frakfam=\sevenfrak \scriptscriptfont\frakfam=\fivefrak
	
\fi
\catcode`\@=\active
\newfam\msbfam
\batchmode\font\twelvemsb=msbm10 scaled\magstep1 \errorstopmode
\ifx\twelvemsb\nullfont\def\Bbb{\bf}
        
	\font\eightbbb=cmb10 at 8pt
	\message{Blackboard bold not available. Replacing with boldface.}
\else   \catcode`\@=11
        \font\tenmsb=msbm10 \font\sevenmsb=msbm7 \font\fivemsb=msbm5
        \textfont\msbfam=\tenmsb
        \scriptfont\msbfam=\sevenmsb \scriptscriptfont\msbfam=\fivemsb
        \def\Bbb{\relax\expandafter\Bbb@}
        \def\Bbb@#1{{\Bbb@@{#1}}}
        \def\Bbb@@#1{\fam\msbfam\relax#1}
        \catcode`\@=\active
	
	\font\eightbbb=msbm8
\fi
        \font\fivemi=cmmi5
        \font\sixmi=cmmi6
        \font\eightrm=cmr8              \def\xrm{\eightrm}
        \font\eightbf=cmbx8             \def\xbf{\eightbf}
        \font\eightit=cmti10 at 8pt     \def\xit{\eightit}
                
        \font\eighttt=cmtt8             
        \font\eightcp=cmcsc8
                      \def\xold{\eighti}
        \font\eightmi=cmmi8
                     \def\xbold{\eightib}
        \font\teni=cmmi10               \def\old{\teni}
        \font\tencp=cmcsc10

        \font\twelvecp=cmcsc10 scaled\magstep1
        
        \font\sixrm=cmr6
        \font\fiverm=cmr5

        \font\eightsy=cmsy8
        \font\sixsy=cmsy6
        \font\eightsl=cmsl8
        \font\sixbf=cmbx6

	 at10pt	
	\font\twelvehelvbold=phvb at12pt
	 at14pt
	\font\sixteenhelvbold=phvb at16pt
	 at16pt



\def\xbold{\xbf}
\def\xold{\xrm}


\def\noblackbox{\overfullrule=0pt}
\noblackbox

\def\eightpoint{
\def\rm{\fam0\eightrm}
\textfont0=\eightrm \scriptfont0=\sixrm \scriptscriptfont0=\fiverm
\textfont1=\eightmi  \scriptfont1=\sixmi  \scriptscriptfont1=\fivemi
\textfont2=\eightsy \scriptfont2=\sixsy \scriptscriptfont2=\fivesy
\textfont3=\tenex   \scriptfont3=\tenex \scriptscriptfont3=\tenex
\textfont\itfam=\eightit \def\it{\fam\itfam\eightit}
\textfont\slfam=\eightsl \def\sl{\fam\slfam\eightsl}
\textfont\ttfam=\eighttt \def\tt{\fam\ttfam\eighttt}
\textfont\bffam=\eightbf \scriptfont\bffam=\sixbf 
                         \scriptscriptfont\bffam=\fivebf
                         \def\bf{\fam\bffam\eightbf}
\normalbaselineskip=10pt}

\newtoks\headtext
\headline={\ifnum\pageno=1\hfill\else
	\ifodd\pageno{\eightcp\the\headtext}{ }\dotfill{ }{\old\folio}
	\else{\old\folio}{ }\dotfill{ }{\eightcp\the\headtext}\fi
	\fi}
\def\makeheadline{\vbox to 0pt{\vss\noindent\the\headline\break
\hbox to\hsize{\hfill}}
        \vskip2\baselineskip}
\newcount\infootnote
\infootnote=0
\newcount\footnotecount
\footnotecount=1
\def\foot#1{\infootnote=1
\footnote{${}^{\the\footnotecount}$}{\vtop{\baselineskip=.75\baselineskip
\advance\hsize by
-\parindent{\eightpoint\rm\hskip-\parindent
#1}\hfill\vskip\parskip}}\infootnote=0\global\advance\footnotecount by
1}
\newcount\refcount
\refcount=1
\newwrite\refwrite
\def\oldsize{\ifnum\infootnote=1\xold\else\old\fi}
\def\ref#1#2{
	\def#1{{{\oldsize\the\refcount}}\ifnum\the\refcount=1\immediate\openout\refwrite=\jobname.refs\fi\immediate\write\refwrite{\item{[{\xold\the\refcount}]} 
	#2\hfill\par\vskip-2pt}\xdef#1{{\noexpand\oldsize\the\refcount}}\global\advance\refcount by 1}
	}
\def\refout{\eightpoint\catcode`\@=11
        \xrm\immediate\closeout\refwrite
        \vskip2\baselineskip
        {\noindent\twelvecp References}\hfill\vskip\baselineskip
        \baselineskip=.75\baselineskip
        \input\jobname.refs
        \baselineskip=4\baselineskip \divide\baselineskip by 3
        \catcode`\@=\active\rm}

\def\skipref#1{\hbox to15pt{\phantom{#1}\hfill}\hskip-15pt}

\def\hepth#1{\href{http://xxx.lanl.gov/abs/hep-th/#1}{arXiv:\allowbreak
hep-th\slash{\xold#1}}}

\def\arxiv#1#2{\href{http://arxiv.org/abs/#1.#2}{arXiv:\allowbreak
{\xold#1}.{\xold#2}}} 
 
\def\jhep#1#2#3#4{\href{http://jhep.sissa.it/stdsearch?paper=#2\%28#3\%29#4}{J. High Energy Phys. {\xbold #1#2} ({\xold#3}) {\xold#4}}}

\def\CQG#1#2#3{Class. Quantum Grav. {\xbold#1} ({\xold#2}) {\xold#3}}
\def\FP#1#2#3{Fortsch. Phys. {\xbold#1} ({\xold#2}) {\xold#3}}

\def\IJMPA#1#2#3{Int. J. Mod. Phys. {\xbf A}{\xbold#1} ({\xold#2}) {\xold#3}}

\def\JHEP{\jhep}
\def\JMP#1#2#3{J. Math. Phys. {\xbold#1} ({\xold#2}) {\xold#3}}
\def\JPA#1#2#3{J. Phys. {\xbf A}{\xbold#1} ({\xold#2}) {\xold#3}}

\def\MPLA#1#2#3{Mod. Phys. Lett. {\xbf A}{\xbold#1} ({\xold#2}) {\xold#3}}

\def\NPB#1#2#3{Nucl. Phys. {\xbf B}{\xbold#1} ({\xold#2}) {\xold#3}}

\def\PLB#1#2#3{Phys. Lett. {\xbf B}{\xbold#1} ({\xold#2}) {\xold#3}}
\def\PR#1#2#3{Phys. Rept. {\xbold#1} ({\xold#2}) {\xold#3}}
\def\PRD#1#2#3{Phys. Rev. {\xbf D}{\xbold#1} ({\xold#2}) {\xold#3}}

\def\PTP#1#2#3{Progr. Theor. Phys. {\xbold#1} ({\xold#2}) {\xold#3}}

\newcount\sectioncount
\sectioncount=0
\def\section#1#2{\global\eqcount=0
	\global\subsectioncount=0
        \global\advance\sectioncount by 1
	\ifnum\sectioncount>1
	        \vskip2\baselineskip
	\fi
\noindent{\twelvecp\the\sectioncount. #2}\par\nobreak
       \vskip.5\baselineskip\noindent
        \xdef#1{{\old\the\sectioncount}}}
\newcount\subsectioncount
\def\subsection#1#2{\global\advance\subsectioncount by 1
\vskip.75\baselineskip\noindent\line{\tencp\the\sectioncount.\the\subsectioncount. #2\hfill}\nobreak\vskip.4\baselineskip\nobreak\noindent\xdef#1{{\old\the\sectioncount}.{\old\the\subsectioncount}}}
\def\immediatesubsection#1#2{\global\advance\subsectioncount by 1
\vskip-\baselineskip\noindent
\line{\tencp\the\sectioncount.\the\subsectioncount. #2\hfill}
	\vskip.5\baselineskip\noindent
	\xdef#1{{\old\the\sectioncount}.{\old\the\subsectioncount}}}
\newcount\subsubsectioncount
\def\subsubsection#1#2{\global\advance\subsubsectioncount by 1
\vskip.75\baselineskip\noindent\line{\tencp\the\sectioncount.\the\subsectioncount.\the\subsubsectioncount. #2\hfill}\nobreak\vskip.4\baselineskip\nobreak\noindent\xdef#1{{\old\the\sectioncount}.{\old\the\subsectioncount}.{\old\the\subsubsectioncount}}}
\newcount\appendixcount
\appendixcount=0
\def\appendix#1{\global\eqcount=0
        \global\advance\appendixcount by 1
        \vskip2\baselineskip\noindent
        \ifnum\the\appendixcount=1
        {\twelvecp Appendix A: #1}\par\nobreak
                        \vskip.5\baselineskip\noindent\fi
        \ifnum\the\appendixcount=2
        {\twelvecp Appendix B: #1}\par\nobreak
                        \vskip.5\baselineskip\noindent\fi
        \ifnum\the\appendixcount=3
        {\twelvecp Appendix C: #1}\par\nobreak
                        \vskip.5\baselineskip\noindent\fi}
\def\acknowledgements{\vskip2\baselineskip\noindent
        \underbar{\it Acknowledgements:}\ }
\newcount\eqcount
\eqcount=0
\def\Eqn#1{\global\advance\eqcount by 1
\ifnum\the\sectioncount=0
	\xdef#1{{\noexpand\oldsize\the\eqcount}}
	\eqno({\oldstyle\the\eqcount})
\else
        \ifnum\the\appendixcount=0
\xdef#1{{\noexpand\oldsize\the\sectioncount}.{\noexpand\oldsize\the\eqcount}}
                \eqno({\oldstyle\the\sectioncount}.{\oldstyle\the\eqcount})\fi
        \ifnum\the\appendixcount=1
	        \xdef#1{{\noexpand\oldstyle A}.{\noexpand\oldstyle\the\eqcount}}
                \eqno({\oldstyle A}.{\oldstyle\the\eqcount})\fi
        \ifnum\the\appendixcount=2
	        \xdef#1{{\noexpand\oldstyle B}.{\noexpand\oldstyle\the\eqcount}}
                \eqno({\oldstyle B}.{\oldstyle\the\eqcount})\fi
        \ifnum\the\appendixcount=3
	        \xdef#1{{\noexpand\oldstyle C}.{\noexpand\oldstyle\the\eqcount}}
                \eqno({\oldstyle C}.{\oldstyle\the\eqcount})\fi
\fi}
\def\eqn{\global\advance\eqcount by 1
\ifnum\the\sectioncount=0
	\eqno({\oldstyle\the\eqcount})
\else
        \ifnum\the\appendixcount=0
                \eqno({\oldstyle\the\sectioncount}.{\oldstyle\the\eqcount})\fi
        \ifnum\the\appendixcount=1
                \eqno({\oldstyle A}.{\oldstyle\the\eqcount})\fi
        \ifnum\the\appendixcount=2
                \eqno({\oldstyle B}.{\oldstyle\the\eqcount})\fi
        \ifnum\the\appendixcount=3
                \eqno({\oldstyle C}.{\oldstyle\the\eqcount})\fi
\fi}
\def\multi{\global\advance\eqcount by 1}
\def\multieqn#1{({\oldstyle\the\sectioncount}.{\oldstyle\the\eqcount}\hbox{#1})}
\def\multiEqn#1#2{\xdef#1{{\oldstyle\the\sectioncount}.{\old\the\eqcount}#2}
        ({\oldstyle\the\sectioncount}.{\oldstyle\the\eqcount}\hbox{#2})}
\def\multiEqnAll#1{\xdef#1{{\oldstyle\the\sectioncount}.{\old\the\eqcount}}}
\newcount\tablecount
\tablecount=0
\def\Table#1#2{\global\advance\tablecount by 1
       \xdef#1{\the\tablecount}
       \vskip2\parskip
       \centerline{\it Table \the\tablecount: #2}
       \vskip2\parskip}
\newtoks\url
\def\Href#1#2{\catcode`\#=12\url={#1}\catcode`\#=\active#2}
\def\href#1#2{{#2}}

\parskip=3.5pt plus .3pt minus .3pt
\baselineskip=14pt plus .1pt minus .05pt
\lineskip=.5pt plus .05pt minus .05pt
\lineskiplimit=.5pt
\abovedisplayskip=18pt plus 4pt minus 2pt
\belowdisplayskip=\abovedisplayskip
\hsize=14cm
\vsize=19cm
\hoffset=1.5cm
\voffset=1.8cm
\frenchspacing
\footline={}
\raggedbottom

\newskip\origparindent
\origparindent=\parindent

\def\ss{\scriptstyle}

\def\*{\partial}
\def\punkt{\,\,.}
\def\komma{\,\,,}

\def\={\!=\!}
\def\small#1{{\hbox{$#1$}}}

\def\fraction#1{\small{1\over#1}}
\def\fr{\fraction}
\def\Fraction#1#2{\small{#1\over#2}}
\def\Fr{\Fraction}

\def\eg{{\it e.g.}}

\def\ie{{\it i.e.}}

\def\a{\alpha}

\def\e{\varepsilon}

\def\id{1\hskip-3.5pt 1}

\def\II{I\hskip-.8pt I}

\def\RR{{\Bbb R}}
\def\CC{{\Bbb C}}
\def\HH{{\Bbb H}}




\def\textfrac#1#2{\raise .45ex\hbox{\the\scriptfont0 #1}\nobreak\hskip-1pt/\hskip-1pt\hbox{\the\scriptfont0 #2}}


\def\frac{\Fr}

\def\mathbb{\Bbb}



\def\ms{{\mathstrut}}




\catcode`@=11
\def\openupnormal{\afterassignment\@penupnormal\dimen@=}
\def\@penupnormal{\advance\normallineskip\dimen@
  \advance\normalbaselineskip\dimen@
  \advance\normallineskiplimit\dimen@}
\catcode`@=12

\def\EqMatrix{\let\quad\enspace\openupnormal6pt\matrix}

%
\line{
\epsfysize=18mm
\epsffile{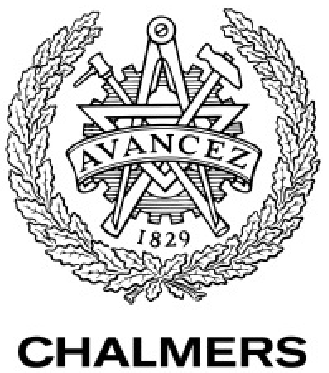}
\hskip5pt
\epsfysize=20mm
\lower5pt\hbox{\epsffile{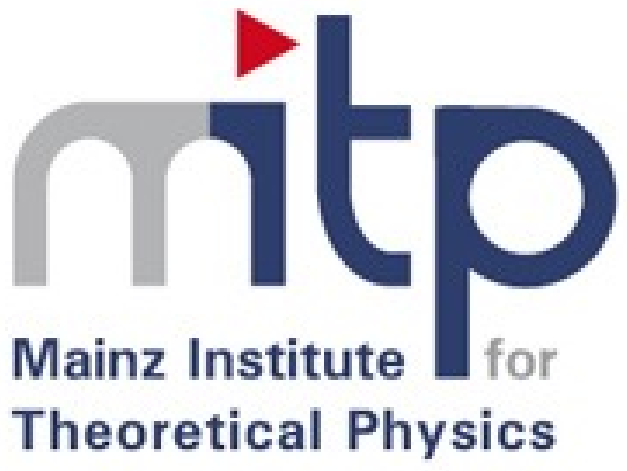}}
\hfill}
\vskip-16mm

\line{\hfill MITP/15-086}
\line{\hfill Gothenburg preprint}
\line{\hfill October, {\old2015}}
\line{\hrulefill}


\headtext={Cederwall: 
``Twistors and supertwistors for exceptional field theory''}

\vfill
\vskip.5cm

\centerline{\sixteenhelvbold
Twistors and supertwistors}

\vskip5mm

\centerline{\sixteenhelvbold
for exceptional field theory}

\vfill

\centerline{\twelvehelvbold Martin Cederwall}

\vfill
\vskip-1cm

\centerline{\it Dept. of Fundamental Physics}
\centerline{\it Chalmers University of Technology}
\centerline{\it SE 412 96 Gothenburg, Sweden}

\vfill

{\narrower\noindent \underbar{Abstract:}
As a means of examining the section condition and its possible
solutions and relaxations, we perform
twistor transforms related to versions of
exceptional field theory with Minkowski signature. The spinor
parametrisation of the momenta naturally 
solves simultaneously both the mass-shell
condition and the (weak) section condition. It is shown that the
incidence relations for multi-particle twistors force them to share a
common section, but not to be orthogonal. The supersymmetric extension
contains additional scalar fermionic variables shown to be
kappa-symmetry invariants. We speculate on some implications,
among them a possible relation to higher spin theory.
\smallskip}
\vfill

\font\xxtt=cmtt6

\vtop{\baselineskip=.6\baselineskip\xxtt
\line{\hrulefill}
\catcode`\@=11
\line{email: martin.cederwall@chalmers.se, arosabal@df.uba.ar\hfill}
\catcode`\@=\active
}

\eject

%

\def\textfrac#1#2{\raise .45ex\hbox{\the\scriptfont0 #1}\nobreak\hskip-1pt/\hskip-1pt\hbox{\the\scriptfont0 #2}}

\def\boxit#1{\vbox{\hrule\hbox{\vrule\kern3pt
             \vbox{\kern3pt#1\kern3pt}\kern3pt\vrule}\hrule}}


\ref\Rosabal{J.A. Rosabal, {\xit ``On the exceptional generalised Lie
derivative for $d\geq7$''}, \arxiv{1410}{8148}.}

\ref\HohmSamtleben{O. Hohm and H. Samtleben, {\xit ``Exceptional field
theory III: $E_{8(8)}$''}, \PRD{90}{2014}{066002} [\arxiv{1406}{3348}].}

\ref\CederwallEdlundKarlsson{M. Cederwall, J. Edlund and A. Karlsson,
  {\xit ``Exceptional geometry and tensor fields''}, 
\jhep{13}{07}{2013}{028} [\arxiv{1302}{6736}].}

\ref\GodazgarGodazgarNicolai{H. Godazgar, M. Godazgar and H. Nicolai,
{\xit ``Einstein-Cartan calculus for exceptional geometry''},
\jhep{14}{06}{2014}{021} [\arxiv{1401}{5984}].}

\ref\BermanCederwallKleinschmidtThompson{D.S. Berman, M. Cederwall,
A. Kleinschmidt and D.C. Thompson, {\xit ``The gauge structure of
generalised diffeomorphisms''}, \jhep{13}{01}{2013}{64} [\arxiv{1208}{5884}].}

\ref\GodazgarGodazgarPerry{H. Godazgar, M. Godazgar and M.J. Perry,
{\xit ``$E_8$ duality and dual gravity''}, \jhep{13}{06}{2013}{044}
[\arxiv{1303}{2035}].} 

\ref\BermanCederwallPerry{D.S. Berman, M. Cederwall and M.J. Perry,
{\xit ``Global aspects of double geometry''}, 
\jhep{14}{09}{2014}{66} [\arxiv{1401}{1311}].}

\ref\BekaertBoulangerHenneaux{X. Bekaert, N. Boulanger, and
M. Henneaux, {\xit ``Consistent deformations of dual formulations of
linearized gravity: a no go result''}, \PRD{67}{2003}{044010}
[\hepth{0210278}].} 

\ref\deWitNicolaiSamtleben{B. de Wit, H. Nicolai and H. Samtleben,
{\xit ``Gauged supergravities, tensor hierarchies, and M-theory''},
\jhep{02}{08}{2008}{044} [\arxiv{0801}{1294}].}

\ref\Duff{M.J. Duff, {\xit ``Duality rotations in string
theory''}, \NPB{335}{1990}{610}.}

\ref\Tseytlin{A.A.~Tseytlin,
  {\xit ``Duality symmetric closed string theory and interacting
  chiral scalars''}, 
  \NPB{350}{1991}{395}.}

\ref\SiegelI{W.~Siegel,
  {\xit ``Two vierbein formalism for string inspired axionic gravity''},
  \PRD{47}{1993}{5453} [\hepth{9302036}].}

\ref\SiegelII{ W.~Siegel,
  {\xit ``Superspace duality in low-energy superstrings''},
  \PRD{48}{1993}{2826} [\hepth{9305073}].}

\ref\SiegelIII{W.~Siegel,
  {\xit ``Manifest duality in low-energy superstrings''},
  in Berkeley 1993, Proceedings, Strings '93 353
  [\hepth{9308133}].}

\ref\HullDoubled{C.M. Hull, {\xit ``Doubled geometry and
T-folds''}, \jhep{07}{07}{2007}{080} [\hepth{0605149}].}

\ref\HullT{C.M. Hull, {\xit ``A geometry for non-geometric string
backgrounds''}, \jhep{05}{10}{2005}{065} [\hepth{0406102}].}

\ref\HullM{C.M. Hull, {\xit ``Generalised geometry for M-theory''},
\jhep{07}{07}{2007}{079} [\hepth{0701203}].}

\ref\HullZwiebachDFT{C. Hull and B. Zwiebach, {\xit ``Double field
theory''}, \jhep{09}{09}{2009}{99} [\arxiv{0904}{4664}].}

\ref\HohmHullZwiebachI{O. Hohm, C.M. Hull and B. Zwiebach, {\xit ``Background
independent action for double field
theory''}, \jhep{10}{07}{2010}{016} [\arxiv{1003}{5027}].}

\ref\HohmHullZwiebachII{O. Hohm, C.M. Hull and B. Zwiebach, {\xit
``Generalized metric formulation of double field theory''},
\jhep{10}{08}{2010}{008} [\arxiv{1006}{4823}].} 

\ref\HohmKwak{O. Hohm and S.K. Kwak, {\xit ``$N=1$ supersymmetric
double field theory''}, \jhep{12}{03}{2012}{080} [\arxiv{1111}{7293}].}

\ref\HohmKwakFrame{O. Hohm and S.K. Kwak, {\xit ``Frame-like geometry
of double field theory''}, \JPA{44}{2011}{085404} [\arxiv{1011}{4101}].}

\ref\JeonLeeParkI{I. Jeon, K. Lee and J.-H. Park, {\xit ``Differential
geometry with a projection: Application to double field theory''},
\jhep{11}{04}{2011}{014} [\arxiv{1011}{1324}].}

\ref\JeonLeeParkII{I. Jeon, K. Lee and J.-H. Park, {\xit ``Stringy
differential geometry, beyond Riemann''}, 
\PRD{84}{2011}{044022} [\arxiv{1105}{6294}].}

\ref\JeonLeeParkIII{I. Jeon, K. Lee and J.-H. Park, {\xit
``Supersymmetric double field theory: stringy reformulation of supergravity''},
\PRD{85}{2012}{081501} [\arxiv{1112}{0069}].}

\ref\HohmZwiebachLarge{O. Hohm and B. Zwiebach, {\xit ``Large gauge
transformations in double field theory''}, \jhep{13}{02}{2013}{075}
[\arxiv{1207}{4198}].} 

\ref\Park{J.-H.~Park,
  {\xit ``Comments on double field theory and diffeomorphisms''},
  \jhep{13}{06}{2013}{098} [\arxiv{1304}{5946}].}

\ref\BermanCederwallPerry{D.S. Berman, M. Cederwall and M.J. Perry,
{\xit ``Global aspects of double geometry''}, 
\jhep{14}{09}{2014}{66} [\arxiv{1401}{1311}].}

\ref\CederwallGeometryBehind{M. Cederwall, {\xit ``The geometry behind
double geometry''}, 
\jhep{14}{09}{2014}{70} [\arxiv{1402}{2513}].}

\ref\HohmLustZwiebach{O. Hohm, D. L\"ust and B. Zwiebach, {\xit ``The
spacetime of double field theory: Review, remarks and outlook''},
\FP{61}{2013}{926} [\arxiv{1309}{2977}].} 

\ref\Papadopoulos{G. Papadopoulos, {\xit ``Seeking the balance:
Patching double and exceptional field theories''}, 
\jhep{14}{10}{2014}{089} [\arxiv{1402}{2586}].}

\ref\HullGlobal{C.M. Hull, 	
{\xit ``Finite gauge transformations and geometry in double field
theory''}, \arxiv{1406}{7794}.}

\ref\PachecoWaldram{P.P. Pacheco and D. Waldram, {\xit ``M-theory,
exceptional generalised geometry and superpotentials''},
\jhep{08}{09}{2008}{123} [\arxiv{0804}{1362}].}

\ref\Hillmann{C. Hillmann, {\xit ``Generalized $E_{7(7)}$ coset
dynamics and $D=11$ supergravity''}, \jhep{09}{03}{2009}{135}
[\arxiv{0901}{1581}].}

\ref\BermanPerryGen{D.S. Berman and M.J. Perry, {\xit ``Generalised
geometry and M-theory''}, \jhep{11}{06}{2011}{074} [\arxiv{1008}{1763}].}    

\ref\BermanGodazgarPerry{D.S. Berman, H. Godazgar and M.J. Perry,
{\xit ``SO(5,5) duality in M-theory and generalized geometry''},
\PLB{700}{2011}{65} [\arxiv{1103}{5733}].} 

\ref\BermanMusaevPerry{D.S. Berman, E.T. Musaev and M.J. Perry,
{\xit ``Boundary terms in generalized geometry and doubled field theory''},
\PLB{706}{2011}{228} [\arxiv{1110}{97}].} 

\ref\BermanGodazgarGodazgarPerry{D.S. Berman, H. Godazgar, M. Godazgar  
and M.J. Perry,
{\xit ``The local symmetries of M-theory and their formulation in
generalised geometry''}, \jhep{12}{01}{2012}{012}
[\arxiv{1110}{3930}].} 

\ref\BermanGodazgarPerryWest{D.S. Berman, H. Godazgar, M.J. Perry and
P. West,
{\xit ``Duality invariant actions and generalised geometry''}, 
\jhep{12}{02}{2012}{108} [\arxiv{1111}{0459}].} 

\ref\CoimbraStricklandWaldram{A. Coimbra, C. Strickland-Constable and
D. Waldram, {\xit ``$E_{d(d)}\times\hbox{\eightbbb R}^+$ generalised geometry,
connections and M theory'' }, \jhep{14}{02}{2014}{054} [\arxiv{1112}{3989}].} 

\ref\CoimbraStricklandWaldramII{A. Coimbra, C. Strickland-Constable and
D. Waldram, {\xit ``Supergravity as generalised geometry II:
$E_{d(d)}\times\hbox{\eightbbb R}^+$ and M theory''}, 
\jhep{14}{03}{2014}{019} [\arxiv{1212}{1586}].}  

\ref\BermanCederwallKleinschmidtThompson{D.S. Berman, M. Cederwall,
A. Kleinschmidt and D.C. Thompson, {\xit ``The gauge structure of
generalised diffeomorphisms''}, \jhep{13}{01}{2013}{64} [\arxiv{1208}{5884}].}

\ref\ParkSuh{J.-H. Park and Y. Suh, {\xit ``U-geometry: SL(5)''},
\jhep{14}{06}{2014}{102} [\arxiv{1302}{1652}].} 

\ref\CederwallII{ M.~Cederwall,
  {\xit ``Non-gravitational exceptional supermultiplets''},
  \jhep{13}{07}{2013}{025} [\arxiv{1302}{6737}].}

\ref\SambtlebenHohmI{O.~Hohm and H.~Samtleben,
  {\xit ``Exceptional field theory I: $E_{6(6)}$ covariant form of
  M-theory and type IIB''}, 
  \PRD{89}{2014}{066016} [\arxiv{1312}{0614}].}

\ref\SambtlebenHohmII{O.~Hohm and H.~Samtleben,
  {\xit ``Exceptional field theory II: $E_{7(7)}$''},
  \PRD{89}{2014}{066016} [\arxiv{1312}{4542}].}

\ref\CederwallUfoldBranes{M. Cederwall, {\xit ``M-branes on U-folds''},
in proceedings of 7th International Workshop ``Supersymmetries and
Quantum Symmetries'' Dubna, 2007 [\arxiv{0712}{4287}].}

\ref\HasslerLust{F. Hassler and D. L\"ust, {\xit ``Consistent
compactification of double field theory on non-geometric flux
backgrounds''}, \jhep{14}{05}{2014}{085} [\arxiv{1401}{5068}].}

\ref\CederwallDuality{M. Cederwall, {\xit ``T-duality and
non-geometric solutions from double geometry''}, \FP{62}{2014}{942}
[\arxiv{1409}{4463}].} 

\ref\AldazabalGranaMarquesRosabal{G. Aldazabal, M. Gra\~na,
D. Marqu\'es and J.A. Rosabal, {\xit ``Extended geometry and gauged
maximal supergravity''}, 
\jhep{13}{06}{2013}{046} [\arxiv{1302}{5419}].}

\ref\AldazabalGranaMarquesRosabalII{G. Aldazabal, M. Gra\~na,
D. Marqu\'es and J.A. Rosabal, {\xit ``The gauge structure of
exceptional field theories and the tensor hierarchy ''}, 
\jhep{14}{04}{2014}{049} [\arxiv{1312}{4549}].}

\ref\PalmkvistHierarchy{J. Palmkvist, {\xit ``Tensor hierarchies,
Borcherds algebras and $E_{11}$''}, \jhep{12}{02}{2012}{066}
[\arxiv{1110}{4892}].} 

\ref\GreitzHowePalmkvist{J. Greitz, P.S. Howe and J. Palmkvist, {\xit ``The
tensor hierarchy simplified''}, \CQG{31}{2014}{087001} [\arxiv{1308}{4972}].} 

\ref\PalmkvistTensor{J. Palmkvist, {\xit ``The tensor hierarchy
algebra''}, \JMP{55}{2014}{011701} [\arxiv{1305}{0018}].}

\ref\HowePalmkvist{P.S. Howe and J. Palmkvist, {\xit ``Forms and
algebras in (half-)maximal supergravity theories''}, 
\hfill\break\arxiv{1503}{00015}.}

\ref\CederwallPalmkvistBorcherds{M. Cederwall and J. Palmkvist, {\xit
``Superalgebras, constraints and partition functions''}, \arxiv{1503}{06215}.}

\ref\BermanBlairMalekPerry{D.S. Berman, C.D.A. Blair, E. Malek and
M.J. Perry, {\xit ``The $O_{D,D}$ geometry of string theory''},
\IJMPA{29}{2014}{1450080} [\arxiv{1303}{6727}].}

\ref\BlairMalek{C.D.A. Blair and E. Malek, {\xit ``Geometry and fluxes
of SL(5) exceptional field theory''}, \jhep{15}{03}{2015}{144} [\arxiv{1412}{0635}].}

\ref\BlumenhagenHasslerLust{R. Blumenhagen, F. Hassler and D. L\"ust,
{\xit ``Double field theory on group manifolds''},
\jhep{15}{02}{2015}{001} [\arxiv{1410}{6374}].}

\ref\BlumenhagenBosqueHasslerLust{R. Blumenhagen, P. du Bosque,
F. Hassler and D. L\"ust, 
{\xit ``Generalized metric formulation of double field theory on group
manifolds''}, \arxiv{1502}{02428}.}

\ref\ParkSuh{J.-H. Park and Y. Suh, {\xit ``U-geometry: SL(5)''}, 
\jhep{14}{06}{2014}{102} [\arxiv{1302}{1652}].}

\ref\KoepsellNicolaiSamtleben{K. Koepsell, H. Nicolai and
H. Samtleben, {\xit ``On the Yangian $[Y(e_8)]$ quantum symmetry of
maximal supergravity in two dimensions''}, \jhep{99}{04}{1999}{023}
[\hepth{9903111}].}

\ref\CoimbraStricklandWaldramTypeII{A. Coimbra, C. Strickland-Constable and
D. Waldram, {\xit ``Supergravity as generalised geometry I: Type II
theories''}, \jhep{11}{11}{2011}{091} [\arxiv{1107}{1733}].} 

\ref\StricklandConstable{C. Strickland-Constable, {\xit ``Subsectors,
Dynkin diagrams and new generalised geometries''}, \arxiv{1310}{4196}.}

\ref\AdSTwistors{M. Cederwall, {\xit ``AdS twistors for higher spin
theory''}, AIP Conf. Proc. {\xbf 767} ({\xold 2005}) {\xold96}
[\hepth{0412222}].} 

\ref\CederwallRosabal{M. Cederwall and J.A. Rosabal, ``$E_8$
geometry'', \jhep{15}{07}{2015}{007}, [\arxiv{1504}{04843}].}

\ref\deWitNicolaiSamtleben{B. de Wit, H. Nicolai and H. Samtleben,
{\xit ``Gauged supergravities, tensor hierarchies, and M-theory''},
\jhep{02}{08}{2008}{044} [\arxiv{0801}{1294}].}

\ref\BossardKleinschmidt{G. Bossard and A. Kleinschmidt, {\xit ``Loops in exceptional field theory''}, \arxiv{1510}{07859}.}

\ref\HohmSamtlebenIII{O. Hohm and H. Samtleben, {\xit ``Exceptional field
theory III: $E_{8(8)}$''}, \PRD{90}{2014}{066002} [\arxiv{1406}{3348}].}

\ref\HohmSiegelZwiebach{O. Hohm, W. Siegel and B. Zwiebach, {\xit
``Doubled $\alpha'$-geometry''}, \jhep{14}{02}{2014}{065} [\arxiv{1306}{2970}].}

\ref\LeeSection{K. Lee, {\xit ``Towards weakly constrained double field
theory''}, \arxiv{1509}{06973}.}

\ref\PenroseTwistor{R.~Penrose and M.A.H.~McCallum,
{\xit ``Twistor theory: An approach to the quantisation of fields and
space-time''}, \PR{6}{1972}{241}, and references therein.}

\ref\Ferber{A. Ferber, {\xit ``Supertwistors and conformal supersymmetry''}, \NPB{132}{1978}{55}.}

\ref\Shirafuji{T. Shirafuji, {\xit ``Lagrangian mechanics of massless particles with spin''}, \PTP{70}{1983}{18}.}

\ref\BengtssonBengtssonCederwall{A.K.H. Bengtsson, I. Bengtsson,
M. Cederwall and N. Linden,, {\xit ``Particles, superparticles and twistors''}, \PRD{36}{1987}{1766}}

\ref\BengtssonCederwall{I. Bengtsson and M. Cederwall, {\xit
``Particles, twistors and the division algebras''}, \NPB{302}{1988}{81}.}

\ref\tentwistorB{N. Berkovits, {\xit ``A supertwistor description of
the massless superparticle in ten-dimensional superspace''}, \PLB{247}{1990}{45}.}

\ref\tentwistorC{M. Cederwall, {\xit ``Octonionic particles and the
$S^7$ symmetry''}, \JMP{33}{1992}{388}.}

\ref\KugoTownsend{T. Kugo and P. Townsend, {\xit ``Supersymmetry and
the division algebras''},\NPB{221}{1983}{357}.} 

\ref\CederwallJordanMech{M.~Cederwall, {\xit ``Jordan algebra
dynamics''}, \PLB{210}{1988}{169}.} 

\ref\AdSTwistorsOne{M. Cederwall, {\xit ``Geometric construction of AdS
twistors''}, \PLB{483}{2000}{257} [\hepth{0002216}].}

\ref\TwistorIntro{M. Cederwall, {\xit ``Introduction to division
algebras, sphere algebras and twistors''}, \hepth{9310115}.}

\ref\BlairMalekPark{C.D.A. Blair, E. Malek and J.-H. Park, {\xit
``M-theory and type IIB from a duality manifest
action''},\jhep{14}{01}{2014}{172} [\arxiv{1311}{5109}].}

\ref\ClausTwistors{P. Claus, M. G\"unayd\i n, R. Kallosh, J. Rahmfeld
and Y. Zunger, {\xit ``Supertwistors as quarks of $SU(2,2|4)$''},
\jhep{99}{05}{1999}{019} [\hepth{9905112}].}

\ref\GelfondVasiliev{O.A. Gelfond and M.A. Vasiliev, {\xit ``Sp(8)
invariant higher spin theory, twistors and geometric BRST formulation
of unfolded field equations''}, \jhep{09}{12}{2009}{021} [\arxiv{0901}{2176}].}

\ref\HatsudaKamimuraSiegel{M. Hatsuda, K. Kamimura and W. Siegel,
{\xit ``Superspace with manifest T-duality from type II
superstring''}, \jhep{14}{06}{2014}{039} [\arxiv{1403}{3887}].}

\ref\NilssonPure{B.E.W.~Nilsson,
{\xit ``Pure spinors as auxiliary fields in the ten-dimensional
supersymmetric Yang--Mills theory''},
\CQG3{1986}{{\xrm L}41}.}

\ref\HowePureI{P.S. Howe, {\xit ``Pure spinor lines in superspace and
ten-dimensional supersymmetric theories''}, \PLB{258}{1991}{141}.}

\ref\HowePureII{P.S. Howe, {\xit ``Pure spinors, function superspaces
and supergravity theories in ten and eleven dimensions''},
\PLB{273}{1991}{90}.} 

\ref\BerkovitsI{N. Berkovits, 
{\xit ``Super-Poincar\'e covariant quantization of the superstring''}, 
\jhep{00}{04}{2000}{018} [\hepth{0001035}].}

\ref\BerkovitsParticle{N. Berkovits, {\xit ``Covariant quantization of
the superparticle using pure spinors''}, \jhep{01}{09}{2001}{016}
[\hepth{0105050}].}

\ref\CederwallNilssonTsimpisI{M. Cederwall, B.E.W. Nilsson and D. Tsimpis,
{\xit ``The structure of maximally supersymmetric super-Yang--Mills
theory --- constraining higher order corrections''},
\jhep{01}{06}{2001}{034} 
[\hepth{0102009}].}

\ref\CederwallNilssonTsimpisII{M. Cederwall, B.E.W. Nilsson and D. Tsimpis,
{\xit ``D=10 super-Yang--Mills at $\ss O(\a'^2)$''},
\JHEP{01}{07}{2001}{042} [\hepth{0104236}].}

\ref\CederwallSuperspace{M. Cederwall, {\xit ``Superspace methods in
string theory, 
supergravity and gauge theory''}, Lectures at the XXXVII Winter School
in Theoretical Physics ``New Developments in Fundamental Interactions
Theories'',  Karpacz, Poland,  Feb. 6-15, 2001, \hepth{0105176}.}

\ref\SpinorialCohomology{M. Cederwall, B.E.W. Nilsson and D. Tsimpis, 
{\xit ``Spinorial cohomology and maximally supersymmetric theories''},
\jhep{02}{02}{2002}{009} [\hepth{0110069}].}

\ref\CederwallBLG{M. Cederwall, {\xit ``N=8 superfield formulation of
the Bagger--Lambert--Gustavsson model''}, \jhep{08}{09}{2008}{116}
[\arxiv{0808}{3242}].}

\ref\CederwallABJM{M. Cederwall, {\xit ``Superfield actions for N=8 
and N=6 conformal theories in three dimensions''},
\jhep{08}{10}{2008}{70}
[\arxiv{0809}{0318}].}

\ref\PureSGI{M. Cederwall, {\xit ``Towards a manifestly supersymmetric
    action for D=11 supergravity''}, \jhep{10}{01}{2010}{117}
    [\arxiv{0912}{1814}].}  

\ref\PureSGII{M. Cederwall, 
{\xit ``D=11 supergravity with manifest supersymmetry''},
    \MPLA{25}{2010}{3201} [\arxiv{1001}{0112}].}

\ref\PureSpinorOverview{M. Cederwall, {\xit ``Pure spinor superfields
--- an overview''}, Springer Proc. Phys. {\xbf153} ({\xrm2013}) {\xrm61} 
[\arxiv{1307}{1762}].}

\ref\Sudbery{A. Sudbery, {\xit ``Division algebras, (pseudo)orthogonal
groups and spinors''}, \xit J.Phys. \xbf A17 \xrm (1984) 939.}

\ref\BandosLukierskiSorokin{I. Bandos, J. Lukierski and D. Sorokin,
{\xit ``Superparticle models with tensorial central charges''},
\PRD{61}{2000}{045002} [\hepth{9904109}].}

\ref\VasilievConformal{M.A. Vasiliev, {\xit ``Conformal higher spin
symmetries of 4d massless supermultiplets and $osp(L,2M)$ invariant
equations in generalized (super)space''}, \PRD{66}{2002}{066006}
[\hepth{0106149}].}


\section\TheSectionCondition{The section condition --- background and motivation}The section condition in doubled geometry
[\Duff\skipref\Tseytlin\skipref\SiegelI\skipref\SiegelII\skipref\SiegelIII\skipref\HullT\skipref\HullDoubled\skipref\HullZwiebachDFT\skipref\HohmHullZwiebachI\skipref\HohmHullZwiebachII\skipref\CoimbraStricklandWaldramTypeII\skipref\HohmKwakFrame\skipref\HohmKwak\skipref\JeonLeeParkI\skipref\JeonLeeParkII\skipref\JeonLeeParkIII\skipref\HohmZwiebachLarge\skipref\Park\skipref\BermanCederwallPerry\skipref\CederwallGeometryBehind\skipref\HohmLustZwiebach\skipref\Papadopoulos\skipref\HullGlobal\skipref\CederwallDuality\skipref\BlumenhagenHasslerLust-\BlumenhagenBosqueHasslerLust]
and exceptional geometry 
[\HullM\skipref\PachecoWaldram\skipref\Hillmann\skipref\BermanPerryGen\skipref\BermanGodazgarPerry\skipref\BermanGodazgarGodazgarPerry\skipref\BermanGodazgarPerryWest\skipref\CoimbraStricklandWaldram\skipref\CoimbraStricklandWaldramII\skipref\BermanCederwallKleinschmidtThompson\skipref\ParkSuh\skipref\CederwallEdlundKarlsson\skipref\CederwallII\skipref\CederwallUfoldBranes\skipref\AldazabalGranaMarquesRosabal\skipref\AldazabalGranaMarquesRosabalII\skipref\SambtlebenHohmI\skipref\SambtlebenHohmII\skipref\HohmSamtlebenIII-\CederwallRosabal]
is the subject of much discussion. 
On the one hand, it is indispensable for the gauge transformations ---
the generalised diffeomorphisms --- to work, and thus it is integral
to a geometric understanding of extended theories. On the other hand,
this also means that not much is known of the geometric principles behind
M-theory when one goes beyond the BPS sector where it is satisfied
(massless modes on top of windings, roughly speaking).
The string-theoretic origin of the double field theory section
condition is well understood, as it is a truncation of the level
matching condition to this sector. No corresponding explanation of the
section condition in exceptional field theory has been proposed.

The section condition is of course what locally makes the extended
theory equivalent to a supergravity theory. 
Seen as a condition on generalised momenta (momenta and winding
charges), it has a r\^ole as a BPS condition. The momenta are constrained to
belong to non-maximal orbits under the ``structure group'' ($O(d,d)$
or $E_{n(n)}$). Applied to a single momentum this is known as the
``weak section condition'', which for a momentum in the module $R_1$ reads 
$$
P^2|_{R_2}=0\punkt\eqn
$$
Here, $R_1$ and $R_2$ are the first two modules in the tensor hierarchy
[\deWitNicolaiSamtleben]. 
In a second-quantised theory, such a constraint does not make sense,
since it does not respect multiplication of fields, and one is led to
the ``strong section condition'', stating that any two derivatives,
acting on any field (or gauge parameter) fulfill the relation
$$
\*\otimes\*|_{R_2}=0\punkt\eqn
$$
This implies that all derivatives belong to a common ``section'', a maximal
vector space of solutions. In a perturbative quantum theory, it is
important not to over-interpret the constraint. A choice of global
section is not allowed, and only stated sharing a common vertex in an
amplitude diagram will obey a relative section condition
[\BossardKleinschmidt]. Non-vanishing amplitudes, and terms in
effective actions, may contain external momenta with no common
section. 
Such a situation seems to go beyond proposals for mild relaxations of
the section condition like the one by Lee [\LeeSection].    
It has indeed been appreciated that higher derivative terms
typically call for modifications of the section condition
[\HohmSiegelZwiebach].

\section\TwistorTransform{The twistor variables}The idea of the
present paper is to find a parametrisation of momenta in terms of
twistor variables
[\PenroseTwistor\skipref\Ferber\skipref\Shirafuji\skipref\BengtssonBengtssonCederwall\skipref\BengtssonCederwall\skipref\tentwistorB\skipref\tentwistorC\skipref\TwistorIntro\skipref\ClausTwistors\skipref\AdSTwistorsOne-\AdSTwistors], which
simultaneously solves the section condition and the mass-shell
condition. 
This turns out to be natural in such a formalism (in fact, we are not aware of a
reasonable way of similarly parametrising solutions to the section
condition only). Indeed, it is only taken together that they carry
meaning as a BPS condition.
One may then consider going off-shell in twistor
space, which typically entails going off the ``spin shell''. 
It is possible that systematic relaxation of the twistor constraints
will lead to a kind of 
higher spin theory. In any case it looks like an interesting way of
investigating the section condition and its possible relaxations.

We will also investigate how locality is implemented in twistor space
through incidence relations, and show that they force multi-particle
twistors to share a common section. They are however not forced to be
orthogonal, which we take as a consistency check of the formalism.
We will give the full details for the model case of $E_{4(4)}\approx
SL(5)$, and indicate more briefly how the transformations work for
$E_{5(5)}\approx Spin(5,5)$ and $E_{6(6)}$.

\subsection\NFourTwistors{$SL(5)$}We now restrict to the structure group
(corresponding to the duality group)
$E_{4(4)}\approx SL(5)$. 
Instead of letting the 4 dimensions of a vector space solution (the
M-theory solution) to the
section condition have
Euclidean signature, as they have when they are internal coordinates
in a compactification, we want Minkowski signature on solutions to
the section condition. This can be achieved by choosing the local
(``Lorentz'') subgroup to be $SO(2,3)$ \foot{This choice is not
unique. Solutions to the weak section condition are elements in the
Grassmannian of 2-planes in 5 dimensions. Vector spaces of
solutions, \ie, solutions to the strong section condition, are planes
intersection along a common line. If this line is time-like, Minkowski
signature is obtained. The same signature may be obtained from
$SO(1,4)$. We prefer the present signature, which allows for real
spinors.}, or, when spinors are included,
its double cover $Spin(2,3)\approx Sp(4,{\Bbb R})$. Also in the type
\II B solution [\BlairMalekPark], 
this real form allows for a section with Minkowski signature.

The momenta are {\it a priori} in the module ${\bf10}$ of $SL(5)$,
which can be written $P_{[mn]}$, or equivalently $P_{[ab]}$ or $P_{(\alpha\beta)}$,
where $m,n=1,\ldots,5$ are fundamental indices of $SL(5)$, and
$a,b=1,\ldots,5$ and $\alpha,\beta=1,\ldots,4$ vector and spinor
indices, respectively, of $Spin(2,3)$. It is assumed that there is some
generalised vielbein (typically flat) 
to convert between curved and flat indices.

The section condition and masslessness condition are
$$
\eqalign{
&P_{[mn}P_{pq]}=0\komma\cr
&P_{ab}P^{ab}=0\punkt\cr
}\eqn
$$
Note that the section condition is $SL(5)$-covariant, while the
on-shell condition requires a generalised metric. In the following, I
will treat them together and use fundamental $Sp(4)$ indices. 
The two conditions are written collectively as
$$
\e^{\gamma\delta}P_{\alpha\gamma}P_{\beta\delta}=0\komma\Eqn\PSquareZero
$$
where the section condition in ${\bf5}$ constitutes the
$\e$-traceless part and $P^2=0$ the $\e$-trace. 
In this one-particle picture, it should be remembered that solutions
to the section condition does {\it not} project down to a
4-dimensional subspace. Rather, we are dealing with the {\it weak}
section condition, whose 
solutions form a real c\^one over the
Grassmannian of 2-planes in 5 dimensions. This is a 7-dimensional
space, and $P^2=0$ brings the dimension down to 6. The dimension of
the space of solutions always equals the dimension of $R_1$ under $E_{n-1}$.

We want a twistor parametrisation 
that solves the constraint (\PSquareZero)
by expressing $P$ as a bilinear in a bosonic spinor. The
dimensionality of the space of solutions tells us immediately that a
single real $\Lambda_\alpha$ is not enough. A pair is the minimum, and we
can put them in a complex spinor $\Lambda_\alpha$. The twistor
parametrisation of the momentum is
$$
P_{\alpha\beta}=\Lambda_{(\alpha}\bar\Lambda_{\beta)}\punkt\Eqn\TwistorP
$$
We now insert this into the constraint on $P$, eq. (\PSquareZero), and
obtain
$$
\e^{\gamma\delta}P_{\alpha\gamma}P_{\beta\delta}
=\e^{\gamma\delta}\Lambda_{(\alpha}\bar\Lambda_{\gamma)}
                  \Lambda_{(\beta}\bar\Lambda_{\delta)}
=-\fr2\e^{\gamma\delta}\Lambda_{[\alpha}\bar\Lambda_{\beta]}
                  \Lambda_{\gamma}\bar\Lambda_{\delta}\punkt
\eqn
$$
In order for the constraints to be satisfied we need a constraint on
$\Lambda$,
$$
\e^{\alpha\beta}\Lambda_{\alpha}\bar\Lambda_{\beta}=0\punkt\Eqn\LambdaConstraint
$$
Considering that the parametrisation of the momentum 
(\TwistorP) also has a $U(1)$
invariance under $\Lambda\rightarrow e^{i\theta}\Lambda$, the 6
degrees of freedom match the ones in $P$ counted earlier.
Note also that the constraint (\LambdaConstraint) on $\Lambda$ is
equally necessary in order to achieve the section condition and the
on-shell constraint, so it seems that they are naturally linked
together in a twistor description. Eq. (\LambdaConstraint) looks
formally identical to the spin-shell constraint obtained from a
massless twistor transform on AdS${}_4$ [\AdSTwistors], which can be
relaxed in order to obtain variables for higher spin theory.  
There, however, the spinors
$\Lambda$ and $\bar\Lambda$ are conjugate to each other, and the
constraint generates the $U(1)$ transformation. Here,
$\Lambda,\bar\Lambda$ describes only momenta, and gives a
configuration space, not a phase space, for the twistors.

Introducing conjugate variables $W^\alpha$ to $\Lambda_\alpha$, the
twistor transform is completed by
$$
W^\alpha=X^{\alpha\beta}\bar\Lambda_\beta\punkt\eqn
$$
From this, we derive the constraint
$$
\Lambda_\alpha W^\alpha-\bar\Lambda_\alpha\bar W^\alpha=0\komma\eqn
$$
which is the generator of the $U(1)$ transformation. It is also clear
that the twistor transform is invariant under $X\rightarrow X+kP$, so
that the choice of base-point $X$ for the world-line is irrelevant.

In twistor space, locality is represented in terms of incidence
relations, some relations that tell us that $(\Lambda,W)$ and $(\Lambda',W')$
correspond to intersecting world-lines, \ie, that the respective
transforms can be written using the same $X$.
We find immediately that 
$$
\Lambda_\alpha W'^\alpha-\bar\Lambda'_\alpha\bar W^\alpha=0\punkt
\Eqn\IncidenceOne
$$
This is not the full answer, though. 
There will be new constraint in the two-particle phase space, obtained
by acting with the generators (\IncidenceOne) on the constraint 
(\LambdaConstraint). This gives a necessary completion of the incidence
relations, namely
$$
\e^{\alpha\beta}\Lambda_\alpha\bar\Lambda'_\beta=0\punkt\Eqn\Necessary
$$
We should now check that the
{\it strong} version of the section condition is satisfied, \ie, that
$$
\e^{\gamma\delta}P_{[\alpha|\gamma|}P'_{\beta]\delta}
-\fr4\e_{\alpha\beta}\e^{\gamma\delta}\e^{\e\varphi}
P_{\gamma\e}P'_{\delta\varphi}
=0\punkt\eqn
$$ 
Here, it is important that the $\e$-trace remains non-vanishing --- we
want generically to have $P\cdot P'\neq0$ for the two momenta, 
only that they lie
in the same linear subspace corresponding to a solution to the strong
section condition, \ie, $P^\ms_{[mn}P'_{pq]}=0$.
Using the constraint (\Necessary)
together with the constraints (\LambdaConstraint) on $\Lambda$ and $\Lambda'$, 
we obtain
$$
\e^{\gamma\delta}P_{[\alpha|\gamma|}P'_{\beta]\delta}
\sim\e^{\gamma\delta}\Lambda_{[\alpha}\Lambda'_\beta
              \bar\Lambda_\gamma\bar\Lambda'_{\delta]}
\sim\e_{\alpha\beta}\e^{\gamma\delta}\Lambda_\gamma\Lambda'_\delta
              \e^{\e\varphi}\bar\Lambda_\e\bar\Lambda'_\varphi
\punkt\eqn
$$
Antisymmetrisation in four indices implies that the expression is pure
$\e$-trace. (Note that the expression vanishes if the primes are
removed.)

We find it very encouraging, indeed a decisive test of the relevance of
the formalism for exceptional geometry, 
that the simplest possible form of incidence
relations, reducing to the constraints on a single twistor for
coinciding spinors, does
precisely what is wanted, namely solving the strong section condition
without yielding orthogonal momenta.

\subsection\NFiveTwistors{$Spin(5,5)$}The twistor transform for the
case of structure group $Spin(5,5)$ will now be described. In order to
have a section with Minkowski signature, the local subgroup is chosen
to be $USp(2,2)\times USp(2,2)$.
Each factor has an invariant antisymmetric tensor $\epsilon_{ab}$ and
a metric $\eta_{a\bar b}$ with signature (2,2).
Then the Lorentz group of the section
is the diagonal subgroup $USp(2,2)\approx Spin(1,4)$. 

The momentum, a
chiral spinor ${\bf16}$ under $Spin(5,5)$, is in the bi-fundamental
$({\bf4},{\bf4})$ under $USp(2,2)\times USp(2,2)$. Even though the
fundamental is complex, the bi-fundamental is pseudo-real, thanks to the
existence of the involution 
$$
v_{aa'}\rightarrow\tilde v_{aa'}=(\sigma(v))_{aa'}=\eta_{a\bar a}\eta_{a'\bar a'}
     \epsilon^{\bar a\bar b}\epsilon^{\bar a'\bar b'}
     \bar v_{\bar b\bar b'}\punkt\eqn
$$
We can choose $\tilde P=P$. 
The weak section condition is in ${\bf10}$ of $Spin(5,5)$, and states
that $P$ is a pure spinor. Together with the condition $P^2=0$, the
constraints read
$$
\epsilon^{a'b'}P_{aa'}P_{bb'}=0\komma\quad
\epsilon^{ab}P_{aa'}P_{bb'}=0\punkt\Eqn\FivePConstraints
$$

The twistor solution of these constraints requires an object in the
fundamental of each component, \ie, $\lambda_a$ and $\mu_{a'}$, with
the momentum formed as
$$
P=\lambda\mu^t+\sigma(\lambda\mu^t)\punkt\Eqn\FiveTransform
$$
This
gives 16 real degrees of freedom. The momentum is invariant under the
$SU(2)\times\RR^+$ transformations 
$$
\eqalign{
(\lambda,\tilde\lambda)&\rightarrow(\lambda,\tilde\lambda)M\komma\cr
(\mu,\tilde\mu)&\rightarrow(\mu,\tilde\mu)(M^{-1})^t\,\cr
}\eqn
$$
where the matrix $M$ is give by
$$
M=\left[\matrix{\alpha&-\bar\beta\cr\beta&\bar\alpha}\right]\komma
\eqn
$$
and where 
$\tilde\lambda_a=\eta_{a\bar a}\epsilon^{\bar a\bar b}\bar\lambda_{\bar b}$ 
and $\alpha,\beta\in\CC$.
In order for the transform (\FiveTransform) to solve the constraints 
(\FivePConstraints), the twistor variables need to satisfy the scalar
constraints
$$
\eta^{a\bar a}\lambda_a\bar\lambda_{\bar a}=0\komma\quad
\eta^{a'\bar a'}\mu_{a'}\bar\mu_{\bar a'}=0\punkt\eqn
$$
The number of on-shell twistor degrees of freedom is $16-4-2=10$,
matching those of the null pure spinor $P$. 
The discussion of incidence relations etc. can be performed in analogy
with the $n=4$ case, and the details will not be given here.

\subsection\NSixTwistors{$E_{6(6)}$}For $n=6$, the structure group is
$E_{6(6)}$. The locally realised group leading to a section with
Minkowski signature is $USp(4,4)$ (with maximal compact subgroup 
$Spin(5)\times Spin(5)$). It is convenient to realise this group as an
orthogonal group over the quaternions, $USp(4,4)\approx
Spin(2,2;\HH)$. Then, as usual [\Sudbery,\KugoTownsend,\BengtssonCederwall],
the $SU(2)$ R-symmetry is realised by right multiplication with unit
quaternions. This is a convenient way of manifesting the
(pseudo-)reality of the fundamental $({\bf8},{\bf2})$, equivalent to an
``$SU(2)$ Majorana condition''.

A momentum in ${\bf27}$ of $E_{6(6)}$ becomes a hermitean and
traceless $(4\times4)$-matrix
with entries in $\HH$. The constraints on $P$ (the section condition
together with ``$P^2=0$'') 
then simply read
$$
P^2=0\komma\Eqn\SixPConstraint
$$
where quaternionic matrix multiplication is implied. The solution
space is 16-dimensional, and consists of null elements in a cone over
the Cayley plane
[\CederwallJordanMech,\BermanCederwallKleinschmidtThompson].

A single ``spinor'' in $({\bf8},{\bf2})$ of $USp(4,4)\times SU(2)$ is
not enough, at least two are needed. The R-symmetry then becomes
$Spin(2,\HH)\approx USp(4)\approx Spin(5)$.
We represent this ``spinor'' $\Lambda$ as a $(4\times2)$-matrix.
This means that a parametrisation
$$
P=\Lambda\Lambda^\dagger\Eqn\SixTransform
$$
will have an invariance under $\Lambda\rightarrow\Lambda M$, where $M$
is a matrix in $Spin(2;\HH$), \ie, $MM^\dagger=\id$ [\Sudbery].
This takes away 10 degrees of freedom. In order for the constraint
(\SixPConstraint) on $P$ to be satisfied, $\Lambda$ has to obey the 6
constraints (in a hermitean $(2\times2)$-matrix)
$$
\Lambda^\dagger\Lambda=0\punkt\Eqn\SixLConstraint
$$
Strictly speaking, the trace should have been subtracted in
eq. (\SixTransform), but it already vanishes due to
eq. (\SixLConstraint).
The counting of the twistor degrees of freedom now gives
$32-10-6=16$, matching the ones in $P$.

\section\SuperTwistors{Supertwistors}It is quite straightforward to
extend the construction to supersymmetric particles. The fermionic
variables in the supertwistor will arise as invariants under
$\kappa$-symmetry. It is therefore desirable to start from an action
to be able to keep proper track of the local symmetries, especially
$\kappa$-symmetry. The alternative would be to perform the
supersymmetric extension more {\it ad hoc} in the twistor formalism,
which seems less satisfactory. This can of
course also be done for bosonic particles.

The construction will be performed specifically for the $SL(5)$ case,
and for minimal supersymmetry. The superparticle action should depend
only on the combination
$$
\Pi^{\alpha\beta}=\dot
X^{\alpha\beta}+\theta^{(\alpha}\dot\theta^{\beta)}\punkt\eqn
$$
It will eventually equal the momentum. 
It is invariant under the global supersymmetry transformations
$$
\eqalign{
\delta_\epsilon X^{\alpha\beta}&=-\epsilon^{(\alpha}\theta^{\beta)}\komma\cr
\delta_\epsilon\theta^\alpha&=\epsilon^\alpha\punkt\cr
}\eqn
$$
The weak section condition
and the masslessness condition must follow from the action, and are
implemented by the introduction of a set of Lagrange multipliers
$V_{\alpha\beta}$ in an antisymmetric matrix. The action is
$$
S=\fr2\int d\tau V_{\alpha\beta}V_{\gamma\delta}
\Pi^{\alpha\gamma}\Pi^{\beta\delta}\punkt\Eqn\SuperAction
$$
The $V$'s are non-dynamical, and as long as they are assumed to be
non-degenerate, can be gauge fixed to $\epsilon$, using the symmetry
generated by the primary constraint $P_V^{\alpha\beta}\approx0$. All following
equations are given after that gauge fixing.
Clearly, the momentum conjugate to $X^{\alpha\beta}$ is
$P_{\alpha\beta}=\epsilon_{\alpha\gamma}\epsilon_{\beta\delta}\Pi^{\gamma\delta}$,
and the constraints (\PSquareZero) are reproduced --- they are the
equations of motion obtained by variation of the Lagrange multipliers. 
In addition, the momentum $\pi_\alpha$ conjugate to $\theta^\alpha$ is
constrained by
$$
\pi_\alpha-P_{\alpha\beta}\theta^\beta\approx0\punkt\eqn
$$
It is obvious that the momentum, obeying (\PSquareZero), will have
vanishing determinant, so some of the fermionic constraints are first
class, generating $\kappa$-symmetry.
It is easily checked that $P$ has half rank precisely when eq.
(\PSquareZero) is satisfied, leading to half-BPS excitations, and
reducing the dynamics to that of an ordinary superparticle in 4
dimensions.
The (local) $\kappa$-symmetry may also be verified directly in the
action, by inserting
$$
\eqalign{
\delta_\kappa X^{\alpha\beta}&=\kappa^{(\alpha}\theta^{\beta)}\komma\cr
\delta_\kappa\theta^\alpha&=\kappa^\alpha\komma\cr
}\eqn
$$
where $P_{\alpha\beta}\kappa^\beta=0$. Solving this condition with 
$\kappa^\alpha=\Pi^{\alpha\beta}\varrho_\beta$ and inserting the
variations in the action gives a result that vanishes modulo
constraints.

The twistor parametrisation of the bosonic momentum is identical to
the bosonic twistor transform, eq. (\TwistorP). 
The relation between the conjugate twistor variables $W$ 
and the original superspace
variables has to be modified, however. It reads
$$
W^\alpha=X^{\alpha\beta}\bar\Lambda_\beta
      +\theta^\alpha\bar\xi\komma\eqn
$$
where the fermionic
variables are constructed as
$$
\xi=\fr2\Lambda_\alpha\theta^\alpha
\komma\quad\bar\xi=\fr2\bar\Lambda_\alpha\theta^\alpha\punkt\eqn
$$
The fermionic variables are easily shown to be invariant under
$\kappa$-symmetry, precisely thanks to the constraint on $\Lambda$, 
eq. (\LambdaConstraint). They are conjugate to each other, 
$\{\xi,\bar\xi\}=1$ and span the full fermionic phase space.
Global supersymmetry transforms the supertwistor variables according
to
$$
\eqalign{
\delta_\epsilon\Lambda_\alpha&=0\komma\cr
\delta_\epsilon W^\alpha&=\fr2\epsilon^\alpha\bar\xi\komma\cr
\delta_\epsilon\xi&=\fr2\epsilon^\alpha\Lambda_\alpha\punkt\cr
}\eqn
$$

\section\Conclusions{Outlook}We have constructed twistor
transforms for exceptional field theory with structure group
$E_{n(n)}$, $n=4,5,6$. The main idea is that the section condition and
the on-shell condition are natural to treat together.

It is unclear if the series can be continued to
higher $n$ (lower $n$ should be simple), 
but we have so far not been able to perform the
construction for $n=7$. This may be connected to the observation that,
in the range where the construction has been worked out, the number of
real components in an unconstrained twistor $\Lambda$ is
$2^{n-1}$. Already at $n=6$, this number is 32 and the R-symmetry is
$Spin(5)$, which can be  
identified with the rotation group of 5 extra coordinates. For $n=7$,
the size of the module needed seems to go beyond the M-theory spinor
at maximal supersymmetry. 
The corresponding procedure in double field theory is the somewhat
trivial procedure of performing separate twistor transforms in the two
sectors of $ O(1,d-1)\times O(1,d-1)\subset O(d,d)$.

Another limitation is that we have only considered ``internal''
directions, although in Minkowski signature, and left the remaining
$11-n$ directions out of the picture. Including them would modify the
on-shell condition in a way that will also change the twistor
transform.

Supersymmetry and superfields in flat superspace is straightforward
for the $E_{n(n)}$ structure groups. Supermultiplets have been
constructed in component language in a number of papers, \eg\
refs. [\CoimbraStricklandWaldramII,\SambtlebenHohmI,\CederwallII]. Giving
a geometric meaning to exceptional superspace seems more difficult,
although some progress has been made in double supergeometry
[\HatsudaKamimuraSiegel]. A very desirable goal would be an
understanding of the structure corresponding to pure spinors for
ordinary superspace and supergeometry 
[\NilssonPure\skipref\HowePureI\skipref\HowePureII\skipref\BerkovitsI\skipref\BerkovitsParticle\skipref\CederwallNilssonTsimpisI\skipref\CederwallNilssonTsimpisII\skipref\CederwallSuperspace-\SpinorialCohomology]. 
It seems that precisely the
section condition stands in the way, and needs to be better understood
for this goal to be 
attained. If at some point the issue is resolved, it should be
possible to construct off-shell supersymmetric actions for extended
supersymmetric field theory and supergeometry along the lines of
refs. 
[\PureSGI\skipref\PureSGII\skipref\CederwallBLG\skipref\CederwallABJM-\PureSpinorOverview].

We do not expect the results to have direct bearing on calculations or
on construction of extended field theories. Rather they may provide an
interesting message for field theory and geometry: that the section
condition ultimately should be taken seriously and arise as equations
of motion, on equal footing with ``$P^2=0$''. 
We do not claim that the na\"\i ve way the weak section
condition is obtained in Section \SuperTwistors\ --- from Lagrange
multipliers in a 
world-line formalism --- has any direct
connection to such a field theory formalism; it is practical rather
than deep. The results may however give some direction concerning
possible relaxation of the section condition, in its weak or strong
version. Going off-shell in the twistor formalism means including an
infinite number of fields with different spin. For ordinary higher
spin theory [\BandosLukierskiSorokin,\VasilievConformal,\GelfondVasiliev], 
this is a natural way to derive a set of variables
(oscillators) for the field theory. Although this applies to AdS
twistors [\ClausTwistors,\AdSTwistorsOne,\AdSTwistors], 
a similar statement could be valid in M-theory, and the present
formalism seems to provide a possible starting point for an
investigation of this issue.

\acknowledgements
The author would like to thank the Mainz Institute for Theoretical
Physics (MITP) for its hospitality and support during the programme
``Stringy geometry'', where part of this work was done.

\refout

\end